\documentclass[graybox]{svmult}

\usepackage{mathptmx}       
\usepackage{helvet}         
\usepackage{courier}        
\usepackage{type1cm}        
\usepackage{makeidx}         
\usepackage{graphicx}        
\usepackage{multicol}        
\usepackage[bottom]{footmisc}

\makeindex             

\def\be{\begin{equation}} 
\def\ee{\end{equation}} 
\def\bea{\begin{eqnarray}} 
\def\eea{\end{eqnarray}} 

\begin{document}

\title*{Partial Differential Equations with Random Noise in Inflationary Cosmology}
\author{Robert H. Brandenberger}
\institute{Robert H. Brandenberger \at Physics Department, McGill University,
Montreal, QC H3A 2T, Canada \email{rhb@physics.mcgill.ca}}
%
%
\maketitle

\abstract*{Random noise arises in many physical problems in which the observer is
not tracking the full system. A case in point is inflationary cosmology, the current
paradigm for  describing the very early universe, where one is often interested only
in the time-dependence of a subsystem. In inflationary cosmology it is assumed
that a slowly rolling scalar field leads to an exponential increase in the size of 
space. At the end of this phase, the scalar
field begins to oscillate and transfers its energy to regular matter. This transfer
typically involves a parametric resonance instability. This article reviews work which 
the author has done in collaboration with Walter Craig studying
the role which random noise can play in the parametric resonance instability
of matter fields in the presence of the oscillatory inflaton field. We find that the
particular idealized form of the noise studied here renders the instability more
effective. As a corollary, we obtain a new proof of Anderson localization.}

\abstract{Random noise arises in many physical problems in which the observer is
not tracking the full system. A case in point is inflationary cosmology, the current
paradigm for  describing the very early universe, where one is often interested only
in the time-dependence of a subsystem. In inflationary cosmology it is assumed
that a slowly rolling scalar field leads to an exponential increase in the size of 
space. At the end of this phase, the scalar
field begins to oscillate and transfers its energy to regular matter. This transfer
typically involves a parametric resonance instability. This article reviews work which 
the author has done in collaboration with Walter Craig studying
the role which random noise can play in the parametric resonance instability
of matter fields in the presence of the oscillatory inflaton field. We find that the
particular idealized form of the noise studied here renders the instability more
effective. As a corollary, we obtain a new proof finiteness of the localization
length in the theory of Anderson localization.}

\section{Background}
\label{sec:1}

This article reviews work done in collaboration with Walter Craig 
applying rigorous results from the theory of random matrix
differential equations to problems motivated by early Universe
cosmology \cite{Craig1, Craig2}. As a corollary, we obtain a new proof
of the positivity of the Lyaponov exponent, corresponding to the
finiteness of the localization length in the theory of Anderson 
localization  \cite{Craig3}.

Over the past two decades, cosmology has developed into a data-driven
field. Thanks to new telescopes we are obtaining high precision data 
about the structure of the universe on large scales. Optical telescopes
are probing the distribution of stellar matter to greater depths, microwave
telescopes have allowed us to make detailed maps of anisotropies in
the cosmic microwave background radiation at fractions of $10^{-5}$
of the mean temperature. In the coming years microwave telescopes
outfitted with polarimeters will allow us to produce polarization maps
of the microwave background, and prototype telescopes are being
developed which will allow us to measure the three-dimensional 
distribution of all baryonic matter (not just the stellar component): this is by 
measuring the redshifted 21cm radiation. 

The data from optical telescopes yield three dimensional maps of the
density distribution of stellar matter in space. This data can be
quantified by taking a Fourier transform of the data and determining
the density power spectrum, the square of the amplitude of the
Fourier modes, as a function of wavenumber $k$. Similarly, the
sky maps of the temperature of the cosmic microwave background
can be quantified by expanding the maps in spherical harmonics and
determining the square of the amplitudes of the coefficients as a
function of the angular quantum number $l$. One of the goals
of modern cosmology is to find a causal mechanism which can explain
the origin of these temperature and density fluctuations.
  
The data are being interpreted in a theoretical framework in which
space-time is a four dimensional pseudo-Riemannian manifold
${\cal{M}}$ with a metric $g_{\mu \nu}$ with signature $(+, -, -, -)$,
and evolves in the presence of matter as determined by the
Einstein field equations
\be \label{EFE}
G_{\mu \nu} \, = \, 8 \pi G T_{\mu \nu} \, ,
\ee
where $G_{\mu \nu}$ is the Einstein tensor constructed from the metric
and its first derivatives, $G$ is Newton's gravitational constant, and
$T_{\mu \nu}$ ss the energy-momentum tensor of matter. 

In physics, it is believed that all fundamental equations of motion
follow from an action principle. The physical trajectories 
extremize the action when considering fluctuations of the fields.
The total action for space-time and matter is
is
\be \label{matter}
S \, = \, \int d^4x \sqrt{-g} \bigl[ \frac{R}{8 \pi G} + {\cal{L}}_m(\varphi_i) \bigr] \, ,
\ee
where $\varphi_i$ are matter fields (functions of space-time which
represent matter), ${\cal{L}}_M$ is the Lagrangian for
the matter fields (which is obtained by covariantizing the matter
action in Special Relativity), $g$ is the determinant of the metric tensor, and $R$
is the Ricci scalar.  For simplicity, cosmologists usually consider
scalar matter fields (and not the fermionic and gauge fields which
represent most of the matter particles which are known to exist in
Nature - the only scalar field known to exist is the Higgs field).

These gravitational field equations (\ref{EFE}) follow from varying the joint
gravitational and matter action $S$ with respect to the metric,
and the equations for matter follow from varying $S$ with respect
to each of the matter fields, leading to
\be \label{matter}
{\cal{D}}^2_g (\varphi_i) \, = \, - \frac{\partial V}{\partial \varphi_i} \, 
\ee
where ${\cal{D}}^2_g$ is the covariant d'Alembertian operator in
the metric $g$, and $V(\varphi )$ is the total potential energy density
of the matter fields. We have assumed above that the kinetic terms
of the matter fields are independent of each other and of canonical
form (the reader not familiar with this physics jargon can simply
take (\ref{matter}) to define what the form of the matter Lagrangian
is).

Cosmologists are lucky since observations show that the metric
of space-time is to a first order homogeneous and isotropic on
large length scales, and hence describable by the metric
\be \label{FRW}
ds^2 \, = \, dt^2 - a(t)^2 \bigl( dx^2 + dy^2 + dz^2 \bigr) \, .
\ee
In the above, $t$ is physical time, and $x, y,$ and $z$ are Cartesian
coordinates on the three-dimensional constant time hypersurfaces.
For simplicity (and because current observations show that this
is an excellent approximation) we have assumed that the spatial
hypersurfaces are spatially flat as opposed to positively curved three
spheres or negatively curved hyperspheres (the three possibilities
for the spatial hypersurfaces consistent with homogeneity and
isotropy). 

The function $a(t)$ is called the ``cosmic scale factor".
In the presence of matter, space-time cannot be static.  
In the absence of external
forces, matter follows geodesics, and matter initially at rest
remains at constant values of $x, y,$ and $z$. Hence, these
coordinates are called ``comoving". The function $a(t)$ thus
represents the spatial radius of a ball of matter locally at
rest. Currently, the Universe is expanding and hence $a(t)$
is an increasing function of time. The Einstein equations (\ref{EFE})
yield the following equations for the scale factor:
\bea
H^2 \, &=& \, \frac{8 \pi G}{3} \rho \,  \label{FRW1} \\
{\dot{\rho}} \, &=& - 3 H (\rho + p) \, , \label{FRW2}
\eea
where $\rho$ and $p$ are the energy density and pressure
density of matter, respectively, and
\be
H(t) \, \equiv \, \frac{{\dot{a}}}{a} 
\ee
is the Hubble expansion rate.

In Standard Big Bang cosmology matter is given as a superposition
of  pressureless ``cold matter" with $p = 0$ and relativistic radiation
with $p = \rho / 3$. At late times, the cold matter dominates and it
then follows from (\ref{FRW1}) and (\ref{FRW2}) that
\be
a(t) \, \sim \, \bigl( \frac{t}{t_0} \bigr)^{2/3} \, ,
\ee
where $t_0$ is the normalization time (often taken to be the
present time). With and without radiation, Standard Big Bang
cosmology suffers from a singularity at $t = 0$. At that
time, the curvature of space-time as well as temperature and
density of matter blow up. This is clearly unphysical: no
physical detector can ever measure an infinite result, and
in addition the assumption that matter can be treated as
an ideal classical fluid breaks down at the high energy densities
when quantum and particle physics effects become important.

In addition, Standard Big Bang cosmology cannot explain the
observed homogeneity and isotropy of the universe, and it
cannot provide a causal mechanism for the generation of
the structure in the universe which current data reveal. The
last point is illustrated in the space-time sketch of Fig. 1.
The vertical axis is time, the horizontal axis gives the physical
dimension of space. The region of causal influence of a point
at the initial time is bounded by the ``horizon", the forward
light cone of the initial point. In Standard Big Bang cosmology the
horizon increases as $t$. In contrast, the physical length of
a particular structure in the universe (which is not gravitationally
bound) grows in proportion to $a(t)$ which in Standard cosmology
grows much more slowly than $t$. Hence, if we trace back
the wavelength $\lambda(t)$ of structures seen at the present time on large
cosmological scales, we see that $\lambda(t) > t$ at early times.
Hence, it is impossible to explain the origin of the seeds which
develop into the structures observed today in a causal way
(since the seeds have to be present in the very early universe).
These problems of Standard Big Bang cosmology motivated the
development of the ``Inflationary Universe" scenario. 

%
\begin{figure}[b]
\sidecaption
\includegraphics[scale=.65]{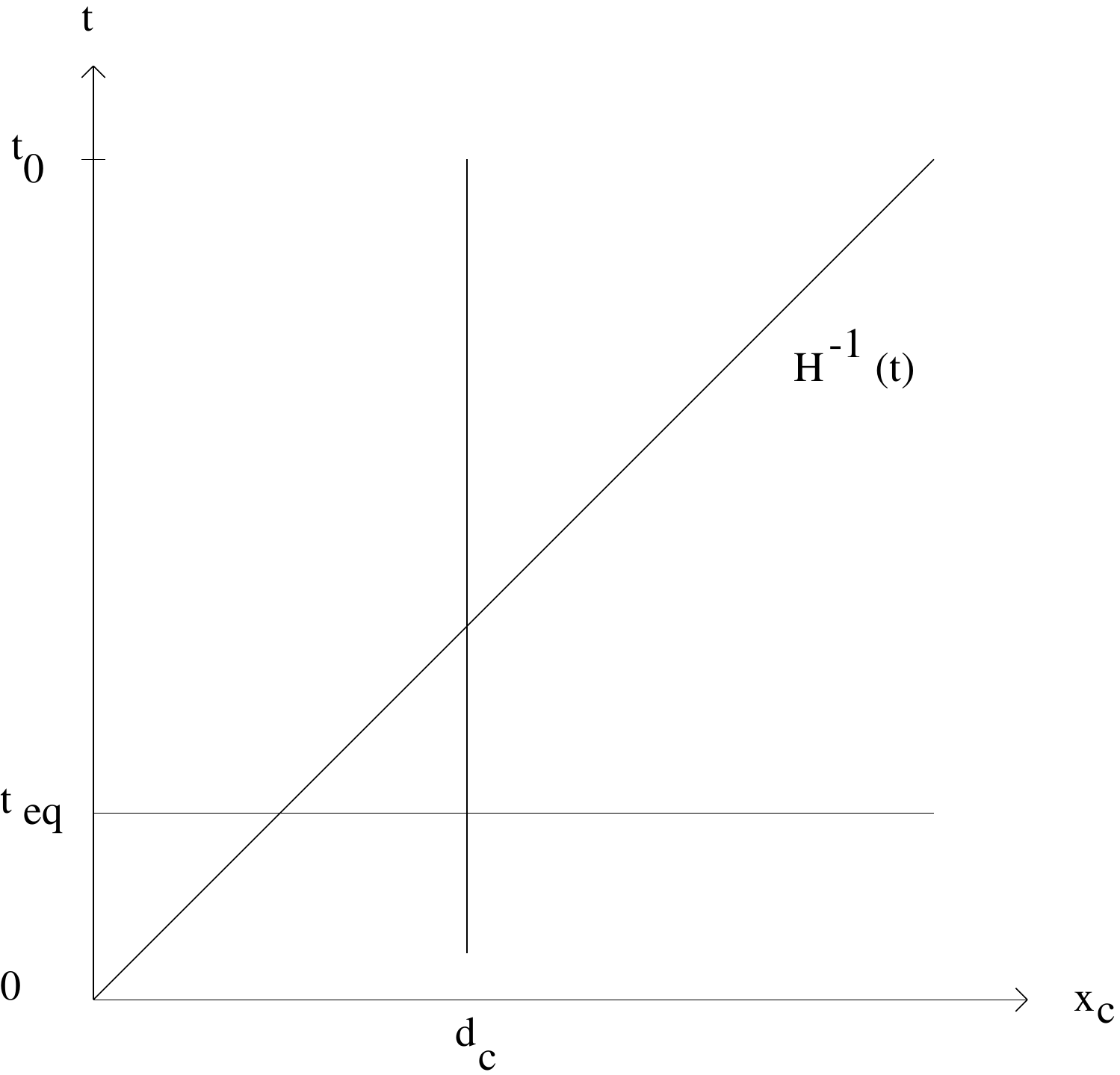}
%
%
\caption{Space-time sketch depicting the ``structure formation problem" of
Standard Big Bang cosmology. The vertical axis is time, the horizontal one
denotes comoving spatial coordinates. The solid vertical line denotes the
wavelength of a mode for which fluctuations are observed, the diagonal curve
through the origin is the cosmological horizon which denotes the limit of
causal influence. As depicted, at early times (in particular at the time $t_{eq}$
when structures can start to grow) the wavelength of the mode
is larger than the horizon and hence no causal structure formation scenario
is possible.}
\label{fig:1}       
\end{figure}

\section{The Inflationary Universe}
\label{sec:2}

The idea behind the Inflationary Universe scenario is very simple \cite{Guth}:
it is postulated that there is an epoch in the very early stage of
cosmology during which the scale factor expands exponentially,
i.e.
\be
a(t) \, \sim \, e^{H t} \, ,
\ee
where here $H$ is a constant. This period lasts from some initial time $t_i$ 
to a final moment $t_R$ (see Fig. 2). 

%
\begin{figure}[b]
\includegraphics[scale=.65]{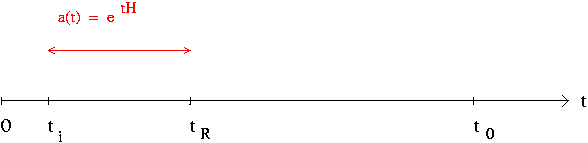}
%
%
\caption{Time line of inflationary cosmology. The period of inflation
begins at the time $t_i$ and ends at $t_R$.}
\label{fig:2}       
\end{figure}

In inflationary cosmology the time evolution of the horizon and
of $\lambda(t)$ are modified compared to what happens in 
Standard Cosmology: the horizon expands exponentially
in the interval between $t_i$ and $t_R$, and so does $\lambda(t)$.
In contrast, the Hubble radius $l_H(t)$ defined as the inverse
Hubble expansion rate
\be
l_H(t) \, = \, H^{-1}(t) \,
\ee
is constant. Provided that the period of inflation is sufficiently
long, then the horizon will at all times be larger than $\lambda(t)$
for any wavelength which can currently be observed (see Fig. 3). Thus,
there is no causality problem to have homogeneity and isotropy
on scales currently observed. As follows from the study of linearized
fluctuations about the background (\ref{FRW}), the
Hubble radius is the upper limit on the length scales on which
fluctuations can be created. From Fig. 3 it can be seen that
in inflationary cosmology perturbation modes originate with a
length smaller than the Hubble radius. Thus, it is possible that
inflation could provide a causal mechanism for the formation
of the structures which are currently observed. In fact, it turns
out the quantum vacuum fluctuations in the exponentially
expanding phase yield such a mechanism \cite{ChibMukh},
but this is not the focus of this article (see \cite{RHBrev}
for a review of  this topic).

\begin{figure}[b]
\includegraphics[scale=.65]{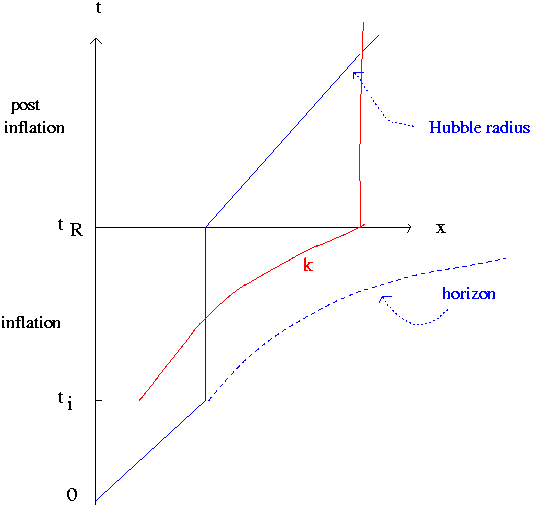}
\caption{Space-time sketch of inflationary cosmology. The vertical
axis denotes time, the horizontal axis is physical distance. During the period
of inflation, the horizon expands exponentially. Similarly, the physical wavelength
of a perturbation mode (the curve denoted by $k$ grows exponentially, and it is 
smaller than the Hubble radius at the beginning of the period of inflation, 
provided the period of inflation lasts sufficiently long.}
\label{fig:3}       
\end{figure}

In order to obtain inflationary expansion in the context of
Einstein's theory of space and time, it follows from Eqs. ({FRW1})
and (\ref{FRW2}) that a form of matter with
\be
p \, = \, - \rho 
\ee
is required. No such equation of state can be obtained using
classical fluids, nor can it be obtained from fields representing
the usual fermionic and gauge degrees of matter, at least in
the context of renormalizable matter theories. Hence, a scalar
field is required to obtain inflation. Even with scalar fields, it
is not easy to obtain inflation since we must ensure that the
potential energy density dominates over kinetic and spatial
gradient energies over a long period of time. This follows 
from the following expressions for the energy density and
pressure of scalar field matter
\bea
\rho \, &=& \, \frac{1}{2} ({\dot{\varphi}})^2 + \frac{1}{2} (\nabla \varphi)^2
+ V(\varphi) \, , \\
p \, &=& \, \frac{1}{2} ({\dot{\varphi}})^2 - \frac{1}{6} (\nabla \varphi)^2
- V(\varphi) \, .
\eea
A typical potential for a scalar field which can lead to inflation
is \cite{Linde}
\be
V(\varphi) \, = \, \frac{1}{2} m^2 \varphi^2 \, .
\ee
The equation of motion for $\varphi$ which follows from (\ref{matter})
is
\be \label{phiEOM}
{\ddot{\varphi}} + 3 H {\dot{\varphi}} - \frac{\nabla^2}{a^2} \varphi \, = \, - V^{'}(\varphi)
\, ,
\ee
where the prime indicates the derivative with respect to $\varphi$.
Inflation can arise if $\varphi$ is slowly rolling, i.e. ${\ddot{\varphi}} \ll H {\dot{\varphi}}$
and ${\dot{\varphi}}^2 \ll V(\varphi)$. Slow rolling is possible for
field values $|\varphi| > m_{pl}$, where $m_{pl}$ is the Planck mass
which is given by
\be
m_{pl}^2 \, = \, \bigl( \frac{8 \pi}{3} G \bigr)^{-1} \, .
\ee 
The slow roll trajectory is given by
\be
\phi(t) \, = \, - \bigl( \frac{2}{9} \bigr)^{1/2} m m_{pl} t \, ,
\ee
and it is in fact a local attractor in initial condition space for large field
values \cite{Kung}.

Once $|\varphi|$ drops below $m_{pl}$, the slow-roll approximation breaks
down, the inflationary period ends, and $\varphi$ begins to oscillate about 
the minimum of its potential at $\varphi = 0$. The amplitude of the oscillations 
is damped by the expansion of space, i.e. by the second term on the left hand
side of (\ref{phiEOM}).

\section{The Reheating Challenge}

The field $\varphi$ which leads to inflation cannot be any of the fields
whose particles we have observed (except possibly the Higgs field if
the latter is non-minimally coupled to gravity \cite{Shaposh}). Any
regular matter (matter which is not $\varphi$) which might have been
present at the beginning of the period of inflation is exponentially
diluted during inflation. Thus, at the end of inflation we have a state
in which no regular matter is present (the regular matter fields are
in their vacuum state), and all energy is locked up in the $\varphi$ field.
Thus, to make inflation into a viable model of the early universe, 
a mechanism is needed to
convert the energy density in $\varphi$ at the end of inflation
into Standard Model particles.

As was discovered in \cite{TB, DK} and worked out later in more
detail in \cite{KLS1, STB, KLS2} (see \cite{ABCM} for a recent review)
the initial energy transfer proceeds via a parametric resonance
instability which is described in more generality by Floquet theory.
Let us represent regular matter by a scalar field $\chi$ which is
weakly coupled to $\varphi$ by an interacting term of the form
\be
{\cal{L}}_{I} \, = \, \frac{1}{2} g \chi^2 \varphi \, ,
\ee
where $g$ is a constant which has dimensions of mass. The
free action for $\chi$ is assumed to be that of a canonical
massless scalar field with no bare potential (i.e. no self-interactions). 
In this case, the equation of motion for $\chi$ becomes
\be
{\ddot{\chi}} + 3 H {\dot{\chi}} - \bigl( \frac{\nabla^2}{a^2} + g \varphi \bigr) \chi 
\, = \, 0 \, .
\ee
Since this is a linear differential equation, each Fourier mode $\chi_k$ 
of $\chi$ will evolve independently according to
\be \label{fund}
{\ddot{\chi}}_k + 3 H {\dot{\chi}}_k + \bigl( \frac{k^2}{a^2} - g \varphi \bigr) \chi_k 
\, = \, 0 \, .
\ee
After the end of the period of inflation, $\varphi$ undergoes damped oscillations.
This leads to a periodic variation of the mass term in (\ref{fund}). This, in turn,
leads to a resonant instability and to energy transfer from $\varphi$ to $\chi$.

Let us for a moment neglect the expansion of space. In this case $H = 0$
and $a = 1$ and then the basic matter equation (\ref{fund}) becomes
\be
 {\ddot{\chi}}_k  + \bigl( k^2 - g {\cal{A}} {\rm{cos}}(\omega t) \bigr) \chi_k 
\, = \, 0 \, .
\ee
where ${\cal{A}}$ is the amplitude of the oscillations of $\varphi$ (constant
if the expansion of space is neglected), and $\omega$ is the frequency
of the oscillations (which equals $m$ in our case). Readers will recognize
this equation as the Mathieu equation \cite{MathieuBook}, an equation 
which has exponentially
growing solutions in resonance bands for $k$ which are centered around
half integer multiples of $\omega$. Because of this instability, there will
be conversion of energy between the $\varphi$ field driving the
resonance and the matter fields, as first pointed out in \cite{TB}. 
This instabiity was later given the name ``preheating" \cite{KLS1}. Since
the instability is exponential, we expect that the time scale of the energy
conversion is small compared to the expansion time $H^{-1}$, and that
hence the approximation of neglecting the expansion of space is
self-consistent.

As discussed in \cite{KLS1, STB, KLS2}, the expansion of space can be
included in an elegant way. In terms of a rescaled field $X_k = a^{3/2} \chi_k$,
the equation of motion (\ref{fund}) becomes an equation of the form
\be
{\ddot{X_k}} + \Omega_k^2(t) X_k \, = \, 0 \, ,
\ee
with an effective frequency $\Omega$ which contains a periodically oscillating
term. An exponential instability persists in this setup, and this justifies
the simplified approach which we focus on in this article, where we 
neglect the expansion of space. In the following, we will extract the
periodic term from the effective frequency, i.e.
\be
\Omega_k^2(t) \, = \, \omega_k^2 + p(\omega t) \, .
\ee

Our starting equation will be
\be \label{Mathieu}
  {\ddot{\chi}}_k  + \bigl( \omega_k^2 + p (\omega t) \bigr) \chi_k 
\, = \, 0 \, .
\ee
where $\omega_k^2$ generalizes the previous setup to the case
in which the $\chi$ field has a non-vanishing mass $m_{\chi}$
\be
\omega_k^2 \, = \, k^2 + m_{\chi}^2 \, ,
\ee
and $p$ is a function with period $2 \pi$. Let us denote the amplitude
of $p$ by $P$. Previous work (see \cite{ABCM} for a review) has shown
that if $P \ll \omega^2$ there is ``narrow-band resonance" (only $k$ modes
within narrow resonance bands experience the instability), whereas if
$P \gg \omega^2$ then there is ``broad-band resonance" in which
all modes with $k \ll \omega$ undergo exponential instability.
In inflationary universe modes with broad-band resonance the
reheating process is very rapid on Hubble time scale.

The Mathieu equation (\ref{Mathieu}) is a special case of a Floquet
type equation. According to Floquet theory \cite{MathieuBook, background}, 
the solutions of (\ref{Mathieu}) scale as
\be
\chi_k(t) \, \sim \, e^{\mu_k t + i \alpha_k t} \, ,
\ee
where the constant $\mu_k + i \alpha_k$ is called the Floquet exponent,
and the real part of it, $\mu_k$, is the Lyapunov exponent. The
constant $\alpha_k$ is the rotation number of the solution, which in the
context of the theory of Schr\"odinger operators is the ``integrated
density of states".

The above setup is, however, too idealized for the purposes of
real cosmology. The field $\varphi$ which yields the inflationary
expansion and the field $\chi$ are only two of many fields. All
of them are excited in the early universe, and they are directly
or indirectly coupled to $\chi$, and they will hence give correction terms
to the basic equation (\ref{Mathieu}). The extra fields are called
the ``environment" in which the system under consideration lives.
The environment is typically describeable by random noise,
which is uncorrelated in time with the time-dependence of
$\varphi(t)$. We will now consider an idealized equation which
includes effects of the noise:
\be \label{Noise}
  {\ddot{\chi}}  + \bigl( - \nabla^2 + p (\omega t) + q(x, t) \bigr) \chi 
\, = \, 0 \, .
\ee
where $q(x, t)$ is a stochastic variable whose time-dependence
is uncorrelated with that of $\varphi(t)$. 

\section{Homogeneous Noise}

Our basic equation (\ref{Noise}) is a second order partial differential 
equation with random coefficients. In phase space, we obtain a random matrix
equation which is first order in time. 

To simplify the analysis, we will first consider the case of homogeneous
noise, i.e. we will assume that the noise function $q$ depends only on
time. In this case, each Fourier mode of $\chi$ continues to evolve
independently and satisfies the equation
\be \label{noise}
  {\ddot{\chi}} _k + \bigl( k^2 + p (\omega t) + q(t) \bigr) \chi 
\, = \, 0 \, .
\ee
This dramatically reduces the mathematical complexity of the problem:
we now have a second order ordinary differential equation rather than
a PDE.

To re-write this equation in the form of a random matrix equation
we introduce the transfer matrix $\Phi_q(t, 0)$ made up of two independent
solutions $\phi_1(t; q)$ and $\phi_2(t; q)$ of (\ref{noise}) and their time 
derivatives:
\[ \Phi_q(t, 0) \, = \, \left( \begin{array}{cc}
\phi_1(t; q) & \phi_2(t; q) \\
{\dot{\phi_1}}(t; q) & {\dot{\phi_2}}(t; q) \end{array} \right).\] 
which satisfies the first order matrix equation
\be
{\dot{\Phi_q}} \, = \, M(q(t), t) \Phi_q \, ,
\ee
where the matrix $M$ is given by
\[ M \, = \, \left( \begin{array}{cc}
0 & 1 \\
- (\omega_k^2 + p + q) & 0 \end{array} \right).\] 
The transfer matrix describes the evolution of the system
from initial time $t = 0$ to final time $t$.

Let us denote the transfer matrix in the absence of noise by
$\Phi_0(t, 0)$. According to Floquet theory
(see e.g \cite{MathieuBook, background}) for mathematical
background), this matrix takes the form
\be
\Phi_0(t, 0) \, = \, P_0(t) e^{Ct} \,
\ee
where $P_0(t)$ is a periodic matrix function with period $\omega^{-1}$,
and $C$ is a constant matrix whose spectrum is
\be
{\rm{spec}}(C) \, = \, \{ \pm \mu(0) \} \, ,
\ee
where $\mu(0)$ is called the Lyapunov exponent in the absence of noise.

To study the effects of noise, we re-write the full transfer matrix
by extracting the transfer matrix in the absence of noise:
\be
\Phi_q(t, 0) \, = \, \Phi_0(t, 0) \Psi_q(t, 0) \, ,
\ee
where the non-triviality of the reduced matrix $\Psi$ describes
the effects of the noise. The reduced transfer matrix satisfies the
equation
\be
{\dot{\Psi}}_q \, = \, S \Psi_q \, ,
\ee
where $S$ is the following matrix:
\[ S \, = \, \Phi_0^{-1} \left( \begin{array}{cc}
0 & 0 \\
- q & 0 \end{array} \right) \Phi_0 \, .\] 

Let $T$ be the period of the oscillation of $\phi$. We can now write the
transfer matrix as a product of transfer matrices over individual oscillation
times:
\be
\Phi_q(NT, 0) \, = \, \prod_{j = 1}^N \Phi_q(jT, (j - 1)T) \, ,
\ee
where $N$ is an integer.

Let us assume that the noise is uncorrelated in time when considered
in different oscillation periods. In addition, let us assume that the noise
is drawn from some probability measure on ${\cal{C}}({\cal{R}})$ such
that $q$ restricted to a period fills a neighborhood of ${\cal{C}}({\cal{R}})$.
In this case, the overall Lyapunov exponent is well defined and can be
extracted using the limit \cite{Craig1}
\be
\mu(q) \, = \, {\rm{lim}}_{N \rightarrow \infty} \frac{1}{NT} 
{\rm{log}} || \prod_{j = 1}^N \Phi_q(JT, (j - 1)T) || \, ,
\ee
where $\| \cdot \|$ indicates a matrix norm. Note that the dependence
on the particular matrix norm vanishes in the limit $N \rightarrow \infty$.

The key result of \cite{Craig1} is that noise which obeys the above-mentioned
conditions renders the instability stronger. More specifically, we have the
following theorem:

\begin{theorem}
Given a random noise function $q(t)$ which is uncorrelated on the time
scale $T$ and which is drawn from a probability measure on ${\cal{C}}({\cal{R}})$
such that $q$ restricted to a period fills a neighborhood of ${\cal{C}}({\cal{R}})$,
then
\be \label{ineq}
\mu(q) \, > \, \mu(0) \, .
\ee
\end{theorem}

Note the strict inequality in the above theorem.
At first sight, this result could be surprising since one might expect that
noise could cut off an instability which occurs in the absence of noise. However,
a physical way to understand the result of the above theorem is to realize
that the noise we have introduced can only add energy to the system rather
than drain energy. Thus, it is consistent to find that noise renders the
resonant instability more effective.

The above theorem follows from the Furstenberg Theorem \cite{Furstenberg}
on random matrices. This theorem takes the following form:

\begin{theorem}
Given a probability distribution $dA$ on $\Psi \in SL(2n, {\cal{R}})$ and defining
$G_A$ as the smallest subgroup of $SL(2n, {\cal{R}})$ containing the support of
$dA$, then if $G_A$ is not compact, and $G_A$ restricted to lines has no
invariant measure, then for almost all independent random sequences
$\{ \Psi_j \}_{j = 1}^{\infty}$ distributed according to $dA$ we have
\be
{\rm{lim}}_{N \rightarrow \infty} \frac{1}{N} 
{\rm{log}}|| \prod_{j = 1}^N \Psi_j || \, = \, \lambda \, > \, 0 \, .
\ee
In addition, for almost all vectors $v_1$ and $v_2$ in ${\cal{R}}^{2n}$ the
exponent $\lambda$ can be extracted via
\be
\lambda \, = \, {\rm{lim}}_{N \rightarrow \infty} \frac{1}{N} 
{\rm{log}} < v_1,  \prod_{j = 1}^N \Psi_j  v_2 > \, . 
\ee
\end{theorem}

Note once again the strict inequality in the above theorem. Applied to our
reheating problem, then for any mode $k$, the above theorems in the
case of $n = 1$ can be used, and they imply that, in the presence of noise, 
the Floquet exponent increases for each value of $k$. In particular, if for
a particular value of $k$ there is no instability in the absence of noise, an
instability will develop in the presence of noise.

The way that our result (\ref{ineq}) follows \footnote{There is, in fact,
a small hole in our proof of Theorem 1: in the case of values of
$k$ in the resonance band of the noiseless system, the 
$\Psi_j$ are not necessarily identically distributed on $SL(2)$ because
of the exponential factor which enters. We still obtain the
rigorous result $\mu(q) > 0$ for all values of $k$, and numerical evidence
confirms the validity of the statement $\mu(q) > \mu(0)$ even
for values of $k$ which are in the resonance band. The application
of our result to Anderson localization involves values of $k$ which are
in the stability bands of the noiseless system and is hence robust - I
thank Walter Craig for pointing out this point.}
from Furtsenberg's Theorem is
the following. Let us take $v_1$ to be an eigenvector of $\Phi_0(T, 0)^t$,
the transverse of the noiseless transfer matrix with eigenvalue $exp(\mu(0) T)$.
Then,
\bea
\mu(q) \, &=& \, {\rm{lim}}_{N \rightarrow \infty} \frac{1}{NT}
{\rm{log}} \bigl( <v_1, \Phi_q(NT, 0) v_2 > \bigr) \nonumber \\
&=& \, {\rm{lim}}_{N \rightarrow \infty} \frac{1}{NT} {\rm{log}} \bigl(
e^{\mu_0 NT} < v_1, \prod_{j = 1}^N \Psi_j(jT, (j - 1)T) v_2 > \bigr) \nonumber \\
&=& \, \mu(0) + \lambda \, > \, \mu(0) \, ,
\eea
where in the last step we have used Furstenberg's Theorem.

From the point of view of physics, the restriction to homogeneous noise
is not realistic. We must allow for inhomogeneous noise functions
$q(x, t)$. This is the topic we turn to in the following section.

\section{Inhomogeneous Noise}

In the case of inhomogeneous noise we must return to the
original partial differential equation (\ref{Noise}) with random
noise. By going to phase space we obtain a first order matrix
operator differential equation
\be
{\dot{\Phi_q}} \, = \, M(q(t, \cdot), t) \Phi_q 
\ee
for the fundamental solution matrix operator $\Phi_q$. 
The Floquet exponent for the inhomogeneous system is
defined by
\be
\mu_q \, = \, {\rm{lim}}_{N \rightarrow \infty} {\rm{log}} || \Phi_q(NT, 0) ||
\, ,
\ee
where, as before, $||$ indicates a norm on the matrix operator space, and
$T$ is the period of the unperturbed system. 

As in the previous section, we will separate out the effects
of the noise by defining
\be
\Phi_q(t) \, = \, \Phi_{q = 0}(t) \Psi_q(t) \, ,
\ee
where $\Phi_{q = 0}$ is the fundamental solution matrix 
in the absence of noise, and $\Psi_q(t)$ is the matrix
which encodes the effects of the random noise. We
wish to compare the value of the Floquet exponent in
the presence of noise with that of the noiseless system.

To dramatically reduce the complexity of the problem we
apply a trick which is commonly used is physics. First, we
introduce an infrared cutoff by replacing the infinite spatial
sections ${\cal{R}}^3$ by a three-dimensional torus of side
length $L$. This renders Fourier space discrete. Secondly, we 
impose an ultraviolet cutoff, namely we eliminate high 
``energy" modes with $k < \Lambda$, where $\Lambda$ is 
the cutoff scale. The fundamental solution matrix of
the cutoff problem is denoted by $\Phi_q^{L, \Lambda}(t)$,
and the corresponding Floquet exponents are also denoted
by superscripts.

After the above steps, our problem can be written in Fourier
space as a ordinary matrix differential equation in ${\cal{R}}^{2n}$,
where $n$ is the number of Fourier modes which are left. 
The first pair of coordinates corresponds to the phase space
coordinates of the first Fourier mode and so forth. In
the absence of noise, the fundamental solution matrix is
block diagonal - there is no mixing between different Fourier
modes. In each block, $\Phi_{q = 0}^{L, \Lambda}$ 
reduces to the transfer matrix of the corresponding Fourier
mode discussed in the previous section. The noise
term $\Psi_q$ introduces mixing between the different
blocks.

Since Furstenberg's Theorem is valid on ${\cal{R}}^{2n}$, the
results of the previous section immediately apply and
we have
\be \label{cutoffresult}
\mu_q^{L, \Lambda} \, > \, \mu_0^{L, \Lambda} \, .
\ee
Note the fact that we have a strict inequality. Note also
that the Floquet exponent in the noiseless case is the
maximum of the Floquet exponents over all values of $k$:
\be 
\mu_0^{L, \Lambda} \, = \, {\rm{max}}_k (\mu_{k, 0}) \, ,
\ee
where $\mu_{k, 0}$ is the Floquet exponent for Fourier mode
$k$ in the absence of noise.

Let us now consider removing the limits, i.e. taking $L \rightarrow \infty$
and $\Lambda \rightarrow \infty$. Since in the absence of noise,
there is no resonance for large $k$ modes, the limit of the right hand
side of (\ref{cutoffresult}) is well defined (in fact, the right hand side
is independent of the cutoffs). For any finite value of the cutoffs,
the result (\ref{cutoffresult}) is true. Hence, the result persists in the
limit when the cutoffs are taken to infinity. However, one loses the
strict inequality sign. Hence, assuming that the limit of the left hand
side of (\ref{cutoffresult}) in fact exists, we obtain our final result
\be
 \mu_q \, \geq \, {\rm{max}}_k (\mu_{k, 0}) \, .
\ee

We in fact expect a stronger result. Let us denote by $\mu_q(k)$ the
Floquet exponent of the dynamical system restricted to the k'th
Fourier mode (the restriction made at the end of the evolution).
Then we expect that due to the mode mixing the maximal
growth rate over all Fourier modes of the noiseless system will
influence all Fourier modes of the system with noise, and that 
hence
\be
 \mu_q(k) \, \geq \, {\rm{max}}_k (\mu_{k, 0}) \, .
\ee
Although we have numerical evidence \cite{Craig2} for the validity
of this result, we have not been able to provide a proof.

\section{New Proof of Anderson Localization}

It is well known that there is a correspondence between
classical time-dependent problems and a time-independent
Schr\"odinger equation. Let us start from the second order
differential equation (\ref{noise}) in the case of homogeneous
noise. Let us now make the following substitutions:
\bea
\chi \, &\rightarrow& \, \psi \nonumber \\
t \, &\rightarrow& \, x  \nonumber \\
\omega_k^2 \, &\rightarrow& \, 2 m E \\
p(\omega t) \, &\rightarrow& \, - 2 m V_p(\omega x) \nonumber \\
q(t) \, &\rightarrow& \, - 2 m V_R(x) \, . \nonumber
\eea
Then, the equation (\ref{noise}) becomes
\be
H \psi \, = \, E \psi \, ,
\ee
with the operator $H$ given by
\be
H \, = \, - \frac{1}{2 m} \frac{\partial^2}{\partial x^2} + V_p(\omega x) + V_q(x) \, ,
\ee
which is the time-independent Schr\"odinger equation for the wavefunction
$\psi$ of an electron of mass $m$ in a periodic potential $V_p(\omega x)$
of period $\omega^{-1}$ in the presence of a random noise term $V_R$ in
the potential.

In the absence of noise there are bands of values of $E$ where there
is no instability, and where hence the wave functions are oscillatory. 
In condensed matter physics, the corresponding
solutions for $\psi$ are known as Bloch wave states.  Theorem 1 now
implies that if a random potential is added, then the solutions for
$\psi$ become unstable. This means that there is one exponentially
growing mode and one exponentially decaying mode. In quantum
mechanics the growing mode is unphysical since it is not normalizable.
Hence, the decaying mode is the only physical mode. This solution
corresponds to a localized wave function. Thus, we have obtained
a new proof of the finiteness of the localization length in the
theory of ``Anderson localization" \cite{Anderson} (for reviews
see e.g. \cite{Revs})

\begin{theorem}
Consider the time-independent Schr\"odinger equation for a
particle in a periodic potential $V_p(\omega x)$, and consider a random
noise contribution $V_q(x)$ which is uncorrelated on the length
scale of the period of $V_p$ and which is drawn from a probability 
measure on ${\cal{C}}({\cal{R}})$ such that $q$ restricted to a period fills 
a neighborhood of ${\cal{C}}({\cal{R}})$. Then the presence of the
noise localizes the wavefunction, and the localization strength is
exponential, i.e. the wavefunction $\psi_q$ in the presence of
noise scales as
\be \label{ineq2}
\psi_q(x) \, \sim  \, exp(- \mu(q) x) \, .
\ee
where $\mu(q)$ is strictly positive on the basis of Theorem 1.
\end{theorem}

Note that our method can only be applied to study Anderson localization
in one spatial dimension.

\section{Conclusions}

We have applied rigorous results from random matrix theory to study
the effects of noise on reheating in inflationary cosmology. We have
found that the type of noise studied here, namely a random noise
contribution to the mass term in the Klein-Gordon equation for
a scalar field representing Standard Model matter, renders the
parametric resonance instability of matter production in the
presence of an oscillating inflaton field more effective. After the
standard duality mapping between a time-dependent classical
field theory problem and a time-independent quantum mechanical
Schr\"odinger problem, we obtain a new proof of the finiteness
of the localization length in the theory of ``Anderson
localization'', a famous result in condensed matter physics. Our
work is an example of how the same rigorous mathematics result 
can find interesting applications to diverse physics problems.

\begin{acknowledgement}

I wish to thank P. Guyenne, D. Nicholls and C. Sulem for organizing this 
conference in honor of Walter Craig, and for inviting me to contribute.
Walter Craig deserves special thanks for collaborating with me on the
topics discussed here, for his friendship over many years, and for
comments on this paper.
The author is supported in part by an NSERC Discovery Grant and by
funds from the Canada Research Chair program.

\end{acknowledgement}


\begin{thebibliography}{99}

\bibitem{Craig1}
V.~Zanchin, A.~ Maia Jr., W.~Craig and R.~Brandenberger,
``Reheating in the presence of noise,''
Phys.\ Rev.\ D {\bf 57}, 4651 (1998)
[arXiv:hep-ph/9709273].

\bibitem{Craig2}
V.~Zanchin, A.~Maia Jr., W.~Craig and R.~Brandenberger,
``Reheating in the presence of inhomogeneous noise,''
Phys.\ Rev.\ D {\bf 60}, 023505 (1999)
[arXiv:hep-ph/9901207].

\bibitem{Craig3}
R.~Brandenberger and W.~Craig,
  ``Towards a New Proof of Anderson Localization,''
  Eur.\ Phys.\ J.\ C {\bf 72}, 1881 (2012)
  [arXiv:0805.4217 [hep-th]].

\bibitem{Guth}
A.~H.~Guth,
  ``The Inflationary Universe: A Possible Solution to the Horizon and Flatness Problems,''
  Phys.\ Rev.\ D {\bf 23}, 347 (1981).

\bibitem{ChibMukh}
V.~F.~Mukhanov and G.~V.~Chibisov,
  ``Quantum Fluctuation and Nonsingular Universe. (In Russian),''
  JETP Lett.\  {\bf 33}, 532 (1981)
  [Pisma Zh.\ Eksp.\ Teor.\ Fiz.\  {\bf 33}, 549 (1981)].

\bibitem{RHBrev}
R.~H.~Brandenberger,
  ``Lectures on the theory of cosmological perturbations,''
  Lect.\ Notes Phys.\  {\bf 646}, 127 (2004)
  [hep-th/0306071].

\bibitem{Linde}
A.~D.~Linde,
  ``Chaotic Inflation,''
  Phys.\ Lett.\ B {\bf 129}, 177 (1983).

\bibitem{Kung}
R.~H.~Brandenberger and J.~H.~Kung,
  ``Chaotic Inflation as an Attractor in Initial Condition Space,''
  Phys.\ Rev.\ D {\bf 42}, 1008 (1990);\\
R.~H.~Brandenberger, H.~Feldman and J.~Kung,
  ``Initial conditions for chaotic inflation,''
  Phys.\ Scripta T {\bf 36}, 64 (1991).

\bibitem{Shaposh}
F.~L.~Bezrukov and M.~Shaposhnikov,
  ``The Standard Model Higgs boson as the inflaton,''
  Phys.\ Lett.\ B {\bf 659}, 703 (2008)
  [arXiv:0710.3755 [hep-th]].

\bibitem{TB}
J.~H.~Traschen and R.~H.~Brandenberger,
  ``Particle Production During Out-of-equilibrium Phase Transitions,''
  Phys.\ Rev.\ D {\bf 42}, 2491 (1990).

\bibitem{DK}
A.~D.~Dolgov and D.~P.~Kirilova,
  ``On Particle Creation By A Time Dependent Scalar Field,''
  Sov.\ J.\ Nucl.\ Phys.\  {\bf 51}, 172 (1990)
  [Yad.\ Fiz.\  {\bf 51}, 273 (1990)].

\bibitem{KLS1}
L.~Kofman, A.~D.~Linde and A.~A.~Starobinsky,
  ``Reheating after inflation,''
  Phys.\ Rev.\ Lett.\  {\bf 73}, 3195 (1994)
  [hep-th/9405187].

\bibitem{STB}
Y.~Shtanov, J.~H.~Traschen and R.~H.~Brandenberger,
  ``Universe reheating after inflation,''
  Phys.\ Rev.\ D {\bf 51}, 5438 (1995)
  [hep-ph/9407247].

\bibitem{KLS2}
L.~Kofman, A.~D.~Linde and A.~A.~Starobinsky,
  ``Towards the theory of reheating after inflation,''
  Phys.\ Rev.\ D {\bf 56}, 3258 (1997)
  [hep-ph/9704452].

\bibitem{ABCM}
R.~Allahverdi, R.~Brandenberger, F.~-Y.~Cyr-Racine and A.~Mazumdar,
  ``Reheating in Inflationary Cosmology: Theory and Applications,''
  Ann.\ Rev.\ Nucl.\ Part.\ Sci.\  {\bf 60}, 27 (2010)
  [arXiv:1001.2600 [hep-th]].

\bibitem{MathieuBook} 
N. McLachlan, “Theory and Applications of Mathieu 
Functions” (Oxford Univ. Press, Clarendon, 1947).
 
\bibitem{background}
R. Carmona and J. Lacroix, “Spectral theory of random Schr¨odinger operators” (1990) Birkh¨auser, Boston. See p. 198 for a 
the proof of existence of the Floquet exponents. 

\bibitem{Furstenberg} L. Pastur and A. Figotin, “Spectra of random and 
almost-periodic operators”, (1991) Springer Verlag, Berlin. See p. 256 for 
the theory of generalized Floquet(Lyapunov) exponents and p. 344 for 
Furstenberg’s theorem.

\bibitem{Anderson} P.W. Anderson, 
``Absence of Diffusion in Certain Random Lattices,''
Phys. Rev. {\bf 109}, 1492 (1958);\\
N.F. Mott and W.D. Twose,
``The Theory of Impurity Conduction,''
Adv. Phys. {\bf 10}, 107 (1961);\\
E. Abrahams, P.W. Anderson, D.C. Licciardello and T.V.
Ramakrishnan,
``Scaling Theory of Localization: Absence of Quantum Diffusion in Two 
Dimensions,''
Phys. Rev. Lett. {\bf 42}, 673 (1970).

\bibitem{Revs} D.J. Thouless,
``Electrons in Disordered Systems and the Theory of Localization,''
Rep. Prog. Phys. {\bf 13}, 93 (1974);\\
P.A. Lee and T.V. Ramakrishnan,
``Disordered Electronic Systems,''
Rep. Mod. Phys. {\bf 57}, 287 (1985);\\
B. Kramer and A. MacKimmon,
``Localization: Theory and Experiment,''
Rep. Prog. Phys. {\bf 56}, 1469 (1993).

\end{thebibliography}
\end{document}